%% file: main.tex
\documentclass{article}
\usepackage{spconf,amsmath,graphicx}
\usepackage{amsmath,amssymb,mathtools,amsfonts}
\usepackage{graphicx,subfig,color,wrapfig,epsfig,epstopdf}
\usepackage{url}
\usepackage{multirow}
\usepackage{listings}
\usepackage{xspace}
\usepackage{cleveref}
\usepackage{INTERSPEECH2019}
\newcommand{\dyce}{{ClusterA}\xspace}
\newcommand{\cs}{{ClusterB}\xspace}
\newcommand{\sync}{{SYNC}\xspace}
\newcommand{\hybrid}{{HYBRID}\xspace}
\newcommand{\adpsgd}{{ADPSGD}\xspace}
\newcommand{\swb}{{SWB}\xspace}
\newcommand{\ch}{{CH}\xspace}

\title{Distributed Deep Learning Strategies For Automatic Speech Recognition}
\name{Wei Zhang, Xiaodong Cui, Ulrich Finkler, Brian Kingsbury, George Saon, David Kung, Michael Picheny}
\address{IBM Research}
\email{\{weiz,cuix,ufinkler,bedk,gsaon,kung,picheny\}@us.ibm.com}
\begin{document}
\ninept
\maketitle
\begin{abstract}
In this paper, we propose and investigate a variety of distributed deep learning strategies for automatic speech recognition (ASR) and evaluate them with a state-of-the-art Long short-term memory (LSTM) acoustic model on the 2000-hour Switchboard (SWB2000), which is one of the most widely used datasets for ASR performance benchmark. We first investigate what are the proper hyper-parameters (e.g., learning rate) to enable the training with sufficiently large batch size without impairing the model accuracy. We then implement various distributed strategies, including Synchronous (\sync) , Asynchronous Decentralized Parallel SGD (ADPSGD) and the hybrid of the two \hybrid, to study their runtime/accuracy trade-off. We show that we can train the LSTM model using \adpsgd in 14 hours with 16 NVIDIA P100 GPUs to reach a 7.6\% WER on the Hub5-2000 Switchboard (SWB) test set and a 13.1\% WER on the CallHome (CH) test set. Furthermore, we can train the model using \hybrid  in 11.5 hours with 32 NVIDIA V100 GPUs without loss in accuracy.

\end{abstract}
\input{intro}
\input{background}

\input{design_impl}

\input{meth}
\input{results}

\vspace{-0.15cm}
\input{future}
\vspace{-0.15cm}
\input{related}
\bibliographystyle{IEEEbib}
\bibliography{refs}

\end{document}

%% file: intro.tex
\section{Introduction}
\label{sec:intro}
\input{dl_am}

Unlike the widely-studied computer vision tasks, such as ImageNet \cite{facebook-1hr,ddl,lars,tencent-imgnet,terngrad}, few studies have been published regarding how to accelerate distributed training for ASR tasks on large public dataset (e.g., SWB2000) with the exception of \cite{msr-1bit, deepspeech2}. Compared to computer vision tasks, such as ImageNet, ASR tasks have very distinct behaviors in terms of distributed computing: (1) The state-of-the-art acoustic models conventionally can only be trained with relatively small batch size (e.g. 256) \cite{saon-interspeech-2017}, unlike ImageNet where a ResNet model can be trained with a batch size of 8192 or larger \cite{facebook-1hr, lars,tencent-imgnet}. (2) Computation/communication ratio is low in ASR tasks. In \Cref{sec:di:large_batch}, we demonstrate that SWB2000 with a state-of-the-art LSTM is five times more challenging to scale out than a ResNet for ImageNet. Therefore, we need to revisit distributed training strategies other than the standard synchronous SGD training for acceleration. In this paper, we attempt to address the abovementioned two issues by (1) increasing the batch size for a high-performance LSTM model without impairing model accuracy (2) using the Asynchronous Decentralized Parallel SGD (ADPSGD) \cite{adpsgd} approach to reduce the communication cost and remove runtime bottlenecks. 

%% file: dl_am.tex
Neural networks with deep architectures have been the dominant acoustic modeling approach for automatic speech recognition (ASR) in recent years. They have yielded state-of-the-art performance as compared to previous technologies based on hidden Markov models (HMMs) and Gaussian mixtures \cite{Hinton_DNNSPM}. In some tasks, Deep Learning (DL) has achieved near human-level ASR performance \cite{saon-interspeech-2017, msr-speech}. It is commonly agreed that the success of DL for ASR relies on the availability of large amount of training data and high-performance computing. Therefore, distributed DL is not only preferred but also necessary in DL ASR to guarantee fast turnaround time for model training. 

%% file: background.tex
\section{background}
\label{sec:background}
\input{dl_speedup}

%% file: dl_speedup.tex
A neural network training algorithm seeks to find a set of parameters $\theta^{\ast}$ that minimizes the discrepancy between the network output $\tilde{Y}$ and the ground truth $Y$.
This is usually accomplished by defining a differentiable cost function $C(\hat{Y},Y)$ and iteratively updating each of the model parameters using some variant of the gradient descent algorithm:{\small
\begin{subequations}\label{eq:gd}
\begin{align}
E_m &= \frac{1}{m}\sum\nolimits_{s=1}^{m}C \left(\hat{Y_s},Y_s\right), \label{eq:gd1} \\
\nabla\theta(t) &= \left({\partial E_m}/ {\partial \theta}\right)(t), \label{eq:gd2} \\
\theta(t+1) &= \theta(t) - \alpha(t)\nabla\theta(t) \label{eq:gd3}
\end{align}
\end{subequations}}
where $\theta(t)$ represents the model parameter at iteration $t$, $\alpha$ is the step size (also known as the learning rate), and $m$ is the batch size.

\begin{figure}[t]
  \centering{
        \includegraphics[width=0.40\textwidth]{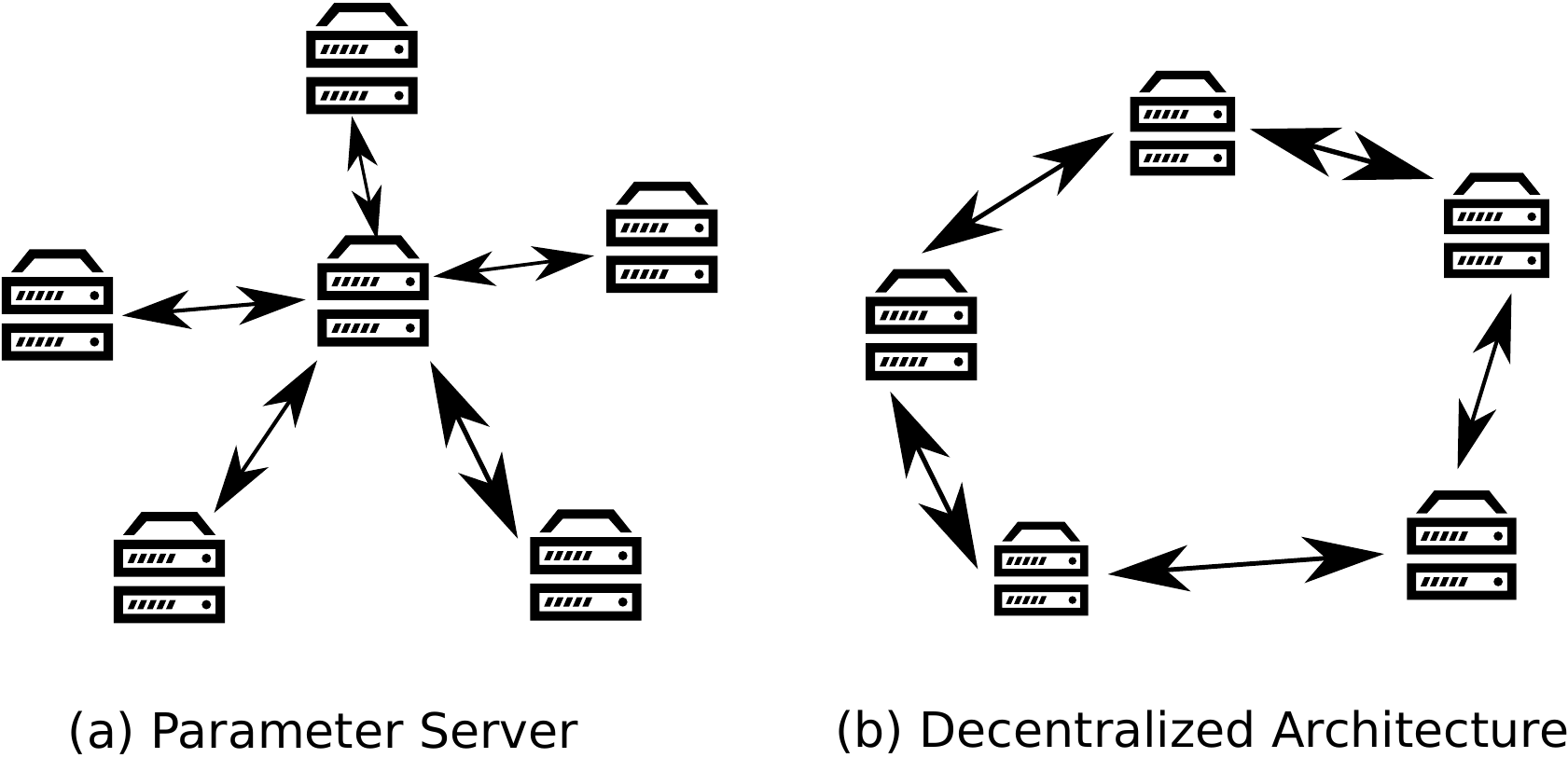}
  }
  \vspace{-1em}
  \caption{A centralized distributed learning architecture and a decentralized distributed learning architecture}
  \label{fig:decen-graphs}
  \vspace{-1em}
\end{figure}

Distributed DL systems, the de facto approach to training large DL tasks, typically adopt a Parameter Server (PS) \cite{distbelief} architecture. \Cref{fig:decen-graphs}(a) depicts a PS architecture in which each learner calculates gradients and transfers them to the PS. The PS then updates the weights and sends them back to the learners. The timestamp, a scalar counter, of the PS's weights is increased by 1 (i.e. from $t$ to $t+1$) for each update. \textit{Staleness}\cite{zhang-ijcai-2016} is defined as the discrepancy between the timestamp of learners' weights which are used to calculate gradients, and the timestamp of the PS's weights. To achieve convergence that is matching to the single learner system, Synchronous SGD (SSGD) is often used. In SSGD, the PS weight update rule is given in \Cref{eqn:update_hard}\footnote{Throughout this paper, we use $\lambda$ to denote the number of learners.}: each learner calculates gradients and receives updated weights in lockstep with the others. The weights used to calculate the gradients are always the same as the one on the PS, thus staleness is 0.
{\footnotesize
\begin{equation}
	\begin{aligned}
		\nabla \theta(t) &= \frac{1}{\lambda}{\sum\nolimits_{i=1}^{\lambda} \nabla\theta_{L_{i}}}, L_{i} \in {L_{1}, ..., L_{\lambda}}\\
		\theta(t+1) &= \theta(t) - \alpha\nabla \theta(t)
	\end{aligned}
 \label{eqn:update_hard}
\end{equation}}
The summation operation in \Cref{eqn:update_hard} is communicative and associative; this is known as the ``Reduce'' operation in the High Performance Computing field. All-Reduction is the operation that reduces (e.g., sums) all the elements and broadcasts the reduction results to each participant. When the message to be ``AllReduced'' is large, as in the DL case, the optimal algorithm maximizes the communication bandwidth utilization by breaking a large message to chunks and pipe-lining the reduction operation with message transferring in a ring topology\cite{fsu-allreduce}. Such an algorithm can finish the AllReduce operation in 2$\times$ M/Bandwidth time, where $M$ is the size of the message, regardless the number of participants. Many such implementations exist, most notably\cite{paddle-paddle,nccl,ddl}.

One key drawback of SSGD is that one slow learner can slowdown the entire training which is known as the \textit{straggler problem}\cite{mapreduce} in distributed computing. To avoid this problem, practitioners proposed Asynchronous SGD (ASGD) which allows each learner to calculate gradients and asynchronously push/pull the gradients/weights to/from PS. The weight update rule in ASGD is given in \Cref{eqn:update_async}:{\footnotesize
\begin{equation}
\begin{aligned}
\nabla \theta(t) &= \nabla\theta_{L_{i}}, L_{i} \in {L_{1}, ..., L_{\lambda}} \\
\theta(t+1) &= \theta(t) - \alpha\nabla \theta(t)
\end{aligned}
\label{eqn:update_async}
\end{equation}}
Staleness in ASGD is proportional to the number of learners in the system\cite{zhang-ijcai-2016, distbelief} and can severely harm convergence\cite{revisit-sync-sgd, zhang2016icdm}. To achieve the best model accuracy,  most distributed deep learning tasks adopt SSGD only\cite{deepspeech2, facebook-1hr,nvidia-lm-scaling,fb-mt-scaling}.




To avoid the straggler problem \textit{\textbf{and}} maintain competitive model accuracy, decentralized distributed computing algorithm\cite{wildfire} is proposed, in both synchronous form Decentralized Parallel SGD (DPSGD)\cite{dpsgd} and in asynchronous form Asynchronous Decentralized Parallel SGD (ADPSGD)\cite{adpsgd}. The architecture of a decentralized SGD system is depicted in \Cref{fig:decen-graphs}(b), where each learner $i$ calculates the gradients, updates its weights, and averages its weights with its neighbor $j$ in a ring topology. DPSGD/ADPSGD weights update rule is defined in \Cref{eqn:update_decentr}. {\footnotesize
\begin{equation}
\begin{aligned}
  \theta(t)'_{L_{i}} &= \theta(t)_{L_{i}} - \alpha\nabla \theta(t)_{L_{i}}\\
  \theta(t+1)_{L_{i}} &= (\theta(t)'_{L_{i}} + \theta(t)'_{L_{j}})/2, L_{i},L_{j} \in {L_{1}, ..., L_{\lambda}}
\end{aligned}
\label{eqn:update_decentr}
\end{equation}}
In DPSGD, assuming the pair-wise weight averaging can be executed multiple times after each gradient calculation, all the learners will reach the same weights, thus the staleness can be zero. In ADPSGD, since computation is overlapped with communication, the staleness is 1 at best. ADPSGD has shown excellent runtime and convergence performance on computer vision tasks with CNN type of models (e.g, ResNet and VGG)\cite{adpsgd}. In this paper, we demonstrate that ADPSGD also achieves excellent runtime and convergence performance on the SWB2000 speech recognition task with an LSTM model.

%% file: design_impl.tex
\section{Design and Implementation}
\label{sec:di}
We describe how to increase the batch size to enable efficient distributed computing for SWB2000-LSTM in \Cref{sec:di:large_batch}. We then describe the design and implementation for different distributed learning strategies in \Cref{sec:di:design}.
\subsection{Increase Batch Size}
\label{sec:di:large_batch}
\begin{figure}[t]
  \centering
  \subfloat[{Computation time (hr) per epoch on one P100 GPU and minimum communication bandwidth requirement (GB/s) under different batch size per learner, SWB2000-LSTM}]{
    \includegraphics[width=0.45\columnwidth, height=3.3cm]{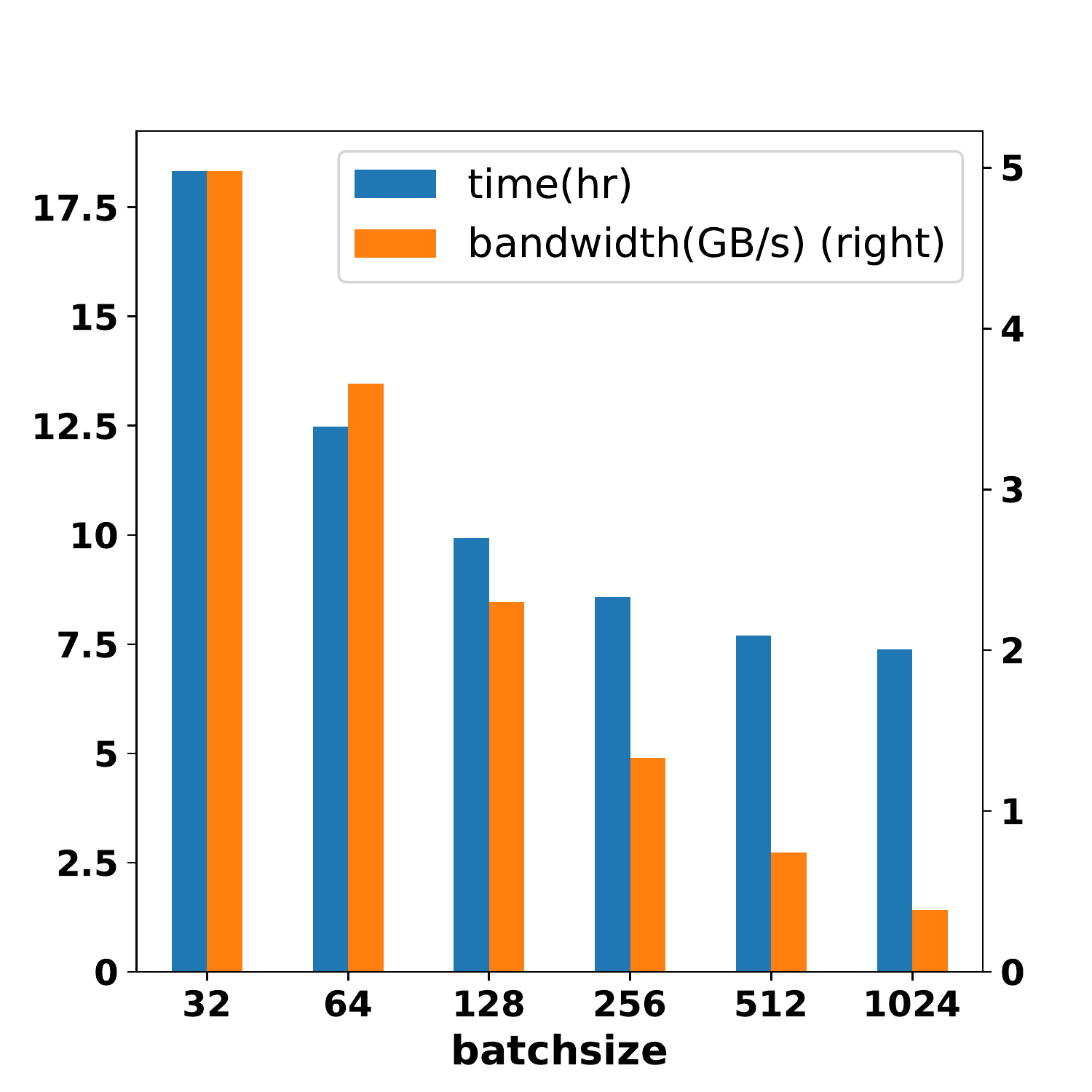}
    \label{fig:single_gpu}
  }
  \hspace{0.3cm}
  \subfloat[{Held-out loss w.r.t epoch for batch size 256 and batch size 2560, SWB2000-LSTM}]{
    \includegraphics[width=0.45\columnwidth, height=3.3cm]{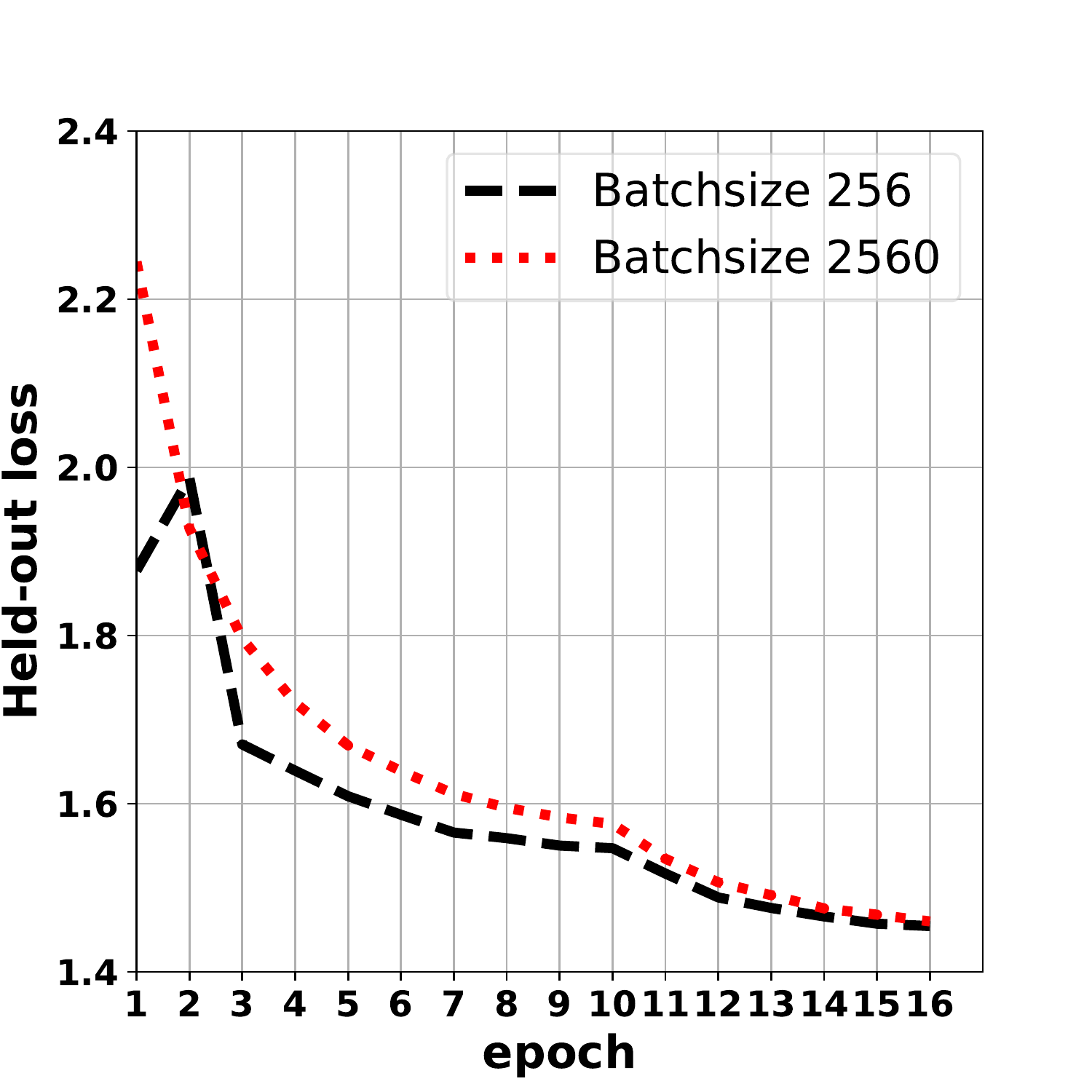}
    \label{fig:large_batch_loss}
  }
  \caption{Computation time / Communication bandwidth requirement and Held-out loss under different batch size, SWB2000-LSTM}    
\end{figure}
 A sufficiently large batch size is necessary for enabling efficient distributed DL for two reasons: (1) The larger the batch, the more learners that can be used. (2) Gradient computation is more efficient with a larger batch size. In \Cref{fig:single_gpu}, blue bars show the computation time per epoch under different batch size for the SWB2000-LSTM\footnote{We describe the details of the LSTM model used in this paper in \Cref{sec:meth:model}} task measured on a P100 GPU. It takes 8.58 hrs to finish one epoch under batch size 256 as compared to 18.33 hrs under batch size 32. Furthermore, with a smaller batch size, more frequent communication is required. In \Cref{fig:single_gpu}, orange bars show the minimum bandwidth requirement to transfer the gradients so that communication time and computation time break even. Batch size 32 per learner requires almost 4X bandwidth (4.98GB/s) as compared with batch size 256 (1.33GB/s).
 Conventional wisdom on SWB2000-LSTM task is batch size larger than 256 significantly lowers the model accuracy\cite{saon-interspeech-2017}. The hyperparameter setup for the batch size 256 configuration is the learning rate is set to be 0.1, momentum is set as 0.9, and learning rate anneals by $\frac{1}{\sqrt2}$ every epoch from the 11$^{th}$ epoch. The training finishes in 16 epochs.
 Inspired by the work proposed in\cite{facebook-1hr}, we are able to increase the batch size from 256 to 2560 without decreasing model accuracy by (1) linearly warming up the base learning rate from 0.1 to 1 in the first 10 epochs and (2) annealing the learning rate by $\frac{1}{\sqrt2}$ from the 11$^{th}$ Epoch. \Cref{fig:large_batch_loss} plots the held-loss w.r.t epochs for batch size 256 and batch size 2560;they are indistinguishable by epoch 16.
 
In the ImageNet-ResNet task, a batch size of 32 takes about 0.18 sec to compute on one P100 GPU, whereas the same batch size for SWB2000-LSTM task takes only 0.067sec to compute. Moreover, the ImageNet-ResNet model size is about 100MB, whereas the SWB2000-LSTM model size is 165MB. \textit{The combination of shorter computation time and larger model size makes SWB2000-LSTM 5x more challenging to parallelize than the ImageNet-ResNet task.}

\subsection{System Design}
\label{sec:di:design}
\newcommand{\allreduce}{AllReduce}
\newcommand{\syncadv}{SyncP2PAdv}
\newcommand{\async}{ADPSGD}
\newcommand{\asyncadv}{AsyncP2PAdv}
Three distributed SGD algorithms are considered and implemented as follows: Synchronous (\sync), Asynchronous Decentralized Parallel SGD (\adpsgd), and the hybrid of these two algorithms \hybrid. 
In \Cref{fig:code-integration}, we sketch our system APIs integration (in bold-font texts) with the underlying DL framework. We assume the underlying DL framework can provide the following functionalities: \textit{g=calGrad(W)}, which calculates the gradients $g$ based on weights $W$ and \textit{W'=apply(W,g)}, which applies gradients $g$ to weights $W$ and returns updated weights $W'$.

\begin{figure}[h]
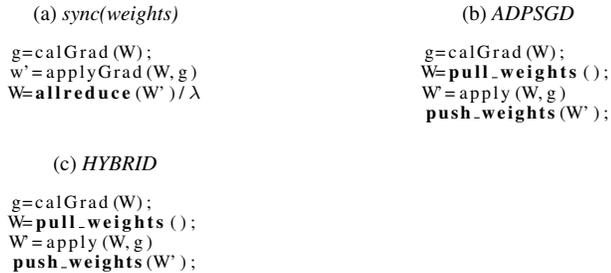

  \caption{Integration of communication protocol with solver.}
  \subfloat[sync(weights)]{\scalebox{1.0}{
\begin{lstlisting}[mathescape=true,emph={allreduce},emphstyle={\textbf},basicstyle=\scriptsize]
g=calGrad(W);
w'=applyGrad(W,g)
W=allreduce(W')/$\lambda$
\end{lstlisting}}\label{fig:allreduce-weights}}
  \hspace{0.3cm}
  \subfloat[\async]{
\begin{lstlisting}[emph={push_weights, pull_weights},emphstyle={\textbf},basicstyle=\scriptsize]
g=calGrad(W);
W=pull_weights();
W'=apply(W,g)
push_weights(W');
  
\end{lstlisting}}
  \hspace{0.3cm}
  \subfloat[\hybrid]{
\begin{lstlisting}[emph={push_weights, pull_weights},emphstyle={\textbf},basicstyle=\scriptsize]
g=calGrad(W);
W=pull_weights();
W'=apply(W,g)
push_weights(W');  
\end{lstlisting}}

  \label{fig:code-integration}
\end{figure}

\textbf{\sync}: 
Summing of the weights and taking their average after every iteration, as shown in \Cref{fig:allreduce-weights}, is equivalent to applying weights update by using the averaged gradients.  
We use the fastest Allreduce implementation available to us (DDL-Allreduce\cite{ddl}) to implement the \sync strategy. As we will show in \Cref{sec:results:runtime}, DDL-Allreduce is 1.2X-3X faster than the open-source MPI\_Allreduce implementation in OpenMPI\cite{openmpi}.  

\begin{figure}[t]
  \centering

    {\includegraphics[width=1.0\columnwidth]{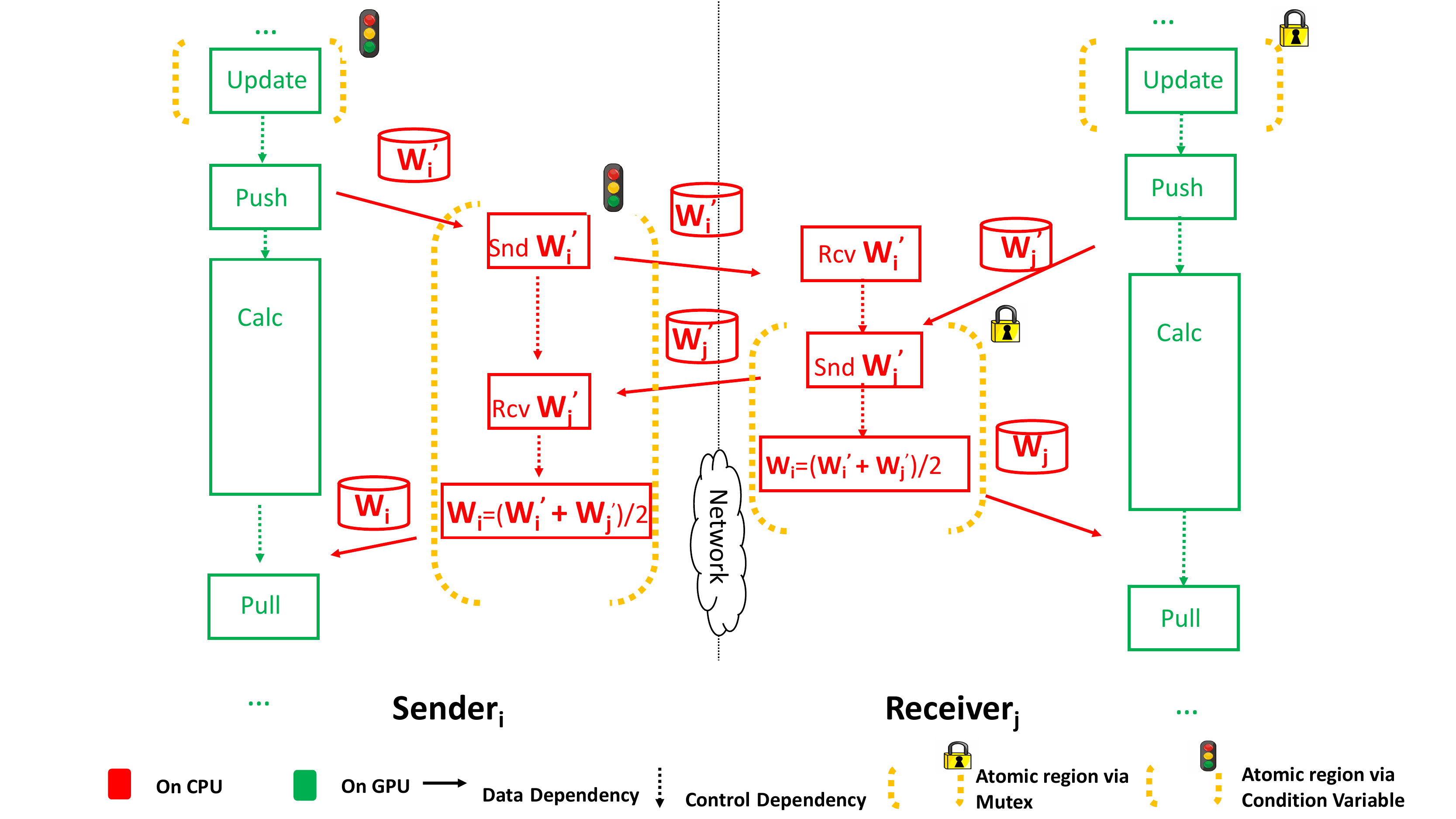}
      \caption{System architecture of ADPSGD.
      }
    \label{fig:arch}
    }
\end{figure}

\textbf{\async}: \Cref{fig:arch} shows the system architecture of \async. Assuming $N$ learners in a system ($N$ is an even number), we designate learners of odd id $i$ ($i \in {1,3,...,N-1}$) as senders and learners of even id $j$ ($j \in {2,4,...,N}$) as receivers. Bipartition of the communication graph guarantees acyclic communication, thus it is deadlock-free. Each sender communicates with its left and right neighbor in alternate iterations. Each sender process runs two threads: a main thread and a communication thread. The main thread calculates the gradients and applies the weight updates. When a new set of weights $W_{i}'$ are generated, the sender signals its communication thread to send $W_{i}'$ to its neighbor learner $j$, receives learner $j$'s weights $W_{j}'$, and then updates its weight $W_{i}$ to be the average of $W_{i}'$ and $W_{j}'$. Additionally, as required in the proof of \cite{adpsgd}, the weight matrix of learners need to be doubly stochastic and symmetric which implies the weight update on GPU and the weight averaging on the CPU must not interleave so that the receiver has a consistent view of weights on the sender. We enforce this atomicity via a condition variable\cite{mesa}. Similarly, each receiver $j$ runs a main thread and a communication thread. In each iteration, a receiver's main thread calculates gradients and then updates its weights in an atomic region.  Meanwhile, the receiver's communication thread waits until the weights are received from its neighbor $i$. Then, the communication thread does the following in an atomic region: (i) sends its weights to its engaging neighbor and updates its weights by averaging its weights with the received weights. The atomicity is enforced via a mutex lock\cite{ArpaciDusseau14-Book}. As compared to the implementation in \cite{adpsgd}, this implementation updates weights on GPU, which runs faster and no gradient information needs be extracted from the underlying solver; the disadvantage of this implementation is that each sending operation is only triggered when new gradients are calculated and there is no way to exchange weights more frequently even when the network is free. Also, GPU weight updates need to wait if the weights are being changed in the communication thread.



\textbf{\hybrid}: Note that the communication threads in \adpsgd essentially runs an Allreduce over learner $i$ and $j$. By replacing the point-to-point message passing with an Allreduce over all learners, we can leverage the optimized fast Allreduce implementation  and also minimize the weights discrepancy among different learners. In essence, \textbf{push\_weights(W')} signals the communication thread to conduct an Allreduce which runs concurrently with gradients calculation, and \textbf{pull\_weights} simply retrieves the Allreduce-d results from the last push and takes an average. The handshaking between communication thread and main thread is a fast lock-free implementation\cite{dyce}. When gradient calculation takes longer time than Allreduce, this scheme should completely overlap communication with computation.

\begin{table}[]
\begin{tabular}{|l|l|l|l|}
\hline
               & \sync & \hybrid & \adpsgd    \\ \hline
Comm/Compute Overlap        & $\times$    & \checkmark      & \checkmark         \\ \hline
Straggler avoidance & $\times$    & $\times$      & \checkmark         \\ \hline
Staleness      & 0    & 1      & at best 1 \\ \hline
\end{tabular}
\caption{Comparison of the runtime and staleness between different algorithms}
\label{tab:protocols_compare}
\vspace{-0.33cm}
\end{table}
\Cref{tab:protocols_compare} summarizes the runtime and staleness comparison between different algorithms. 

%% file: meth.tex
\section{Methodology}
\label{sec:meth}

\subsection{Software and Hardware}
PyTorch 0.5.0 is the underlying DL framework. We use the CUDA 9.2 compiler, the CUDA-aware OpenMPI 3.1.1, and g++ 4.8.5 compiler to build our communication library, which connects with PyTorch via a Python-C interface. 

We develop and experiment our systems on a production-run cluster \dyce, which has 4 servers in total. Each server is equipped with 14-core Intel Xeon E5-2680 v4 2.40GHz processor, 1TB main memory, and 4 P100 GPUs.
We also run a 32-GPU experiment on a 4-server high-GPU-density experimental cluster \cs. Each \cs server is equipped with 18-core Intel Xeon E5-2697 2.3GHz processor, 1TB main memory, and 8 V100 GPUs. Both servers are connected by 100Gbit/s ethernet. On both servers, GPUs and CPUs are connected via PCIe Gen3 bus, which has a 16GB/s peak bandwidth in each direction.



\subsection{DL Models and Dataset}
\label{sec:meth:model}
The acoustic model is an LSTM with 6 bi-directional layers. Each layer contains 1,024 cells (512 cells in each direction). On top of the LSTM layers, there is a linear projection layer with 256 hidden units followed by a softmax output layer with 32,000 units corresponding to context-dependent HMM states. The LSTM is unrolled with 21 frames and trained with non-overlapping feature subsequences of that length.  The feature input is a fusion of FMLLR (40-dim), i-Vector (100-dim), and logmel with its delta and double delta (40-dim $\times$3).

The language model (LM) is rebuilt using publicly available training data, including Switchboard, Fisher, Gigaword, and Broadcast News, and Conversations. Its vocabulary has 85K words and it has 36M 4-grams.

%% file: results.tex
\section{Experimental Results}
\vspace{-0.25cm}
\label{sec:res}
\subsection{Convergence Results}
\vspace{-0.05cm}
\label{sec:res:conv}

\begin{figure}[t]
  \centering
  \subfloat[{Held-out loss comparison between Baseline, \sync, \adpsgd, and \hybrid}]
  {\includegraphics[width=0.45\columnwidth, height=3.3cm]{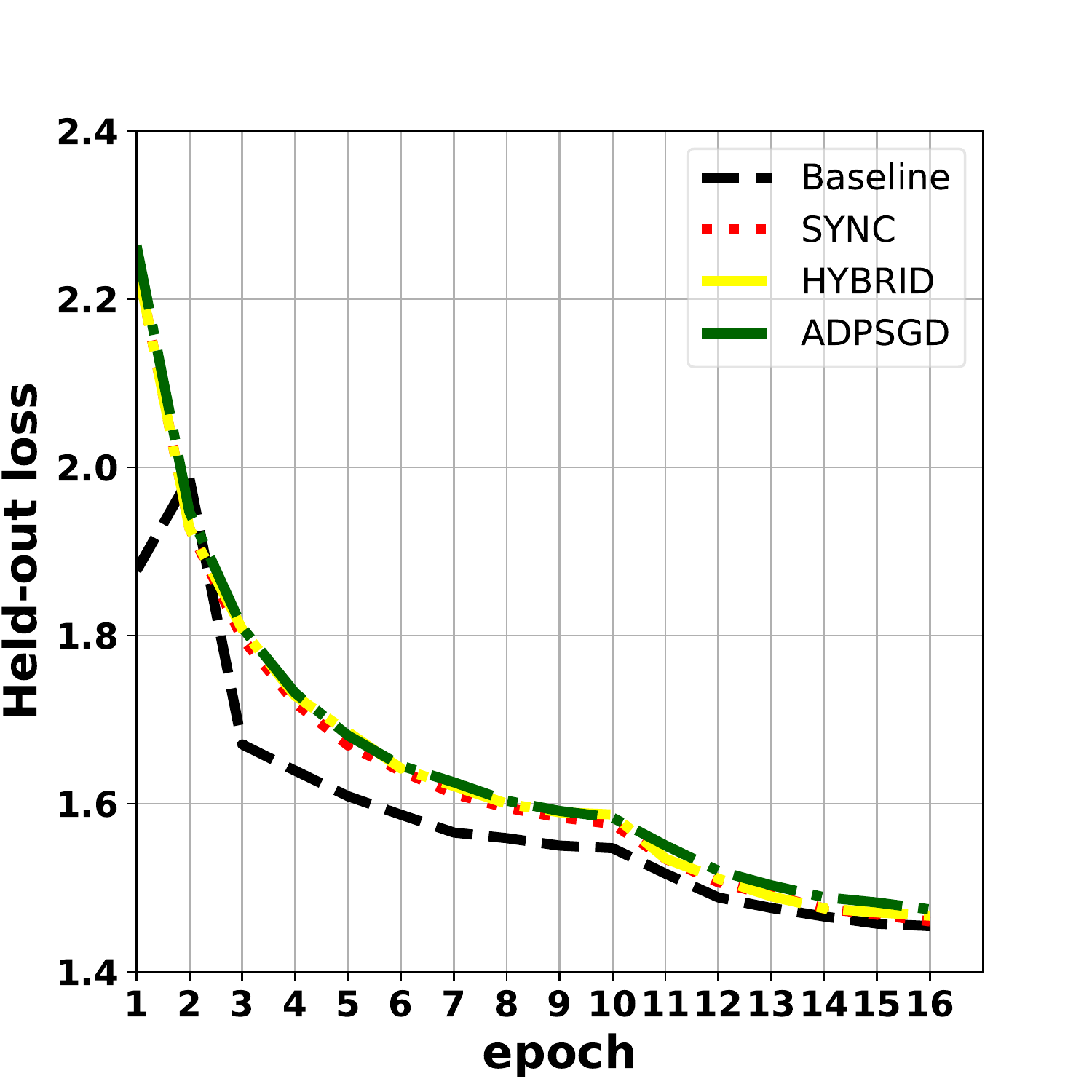}
    \label{fig:loss}
  }
  \hspace{0.2cm}
  \subfloat[{Speed-up comparison between different strategies}]
  {\includegraphics[width=0.45\columnwidth, height=3.3cm]{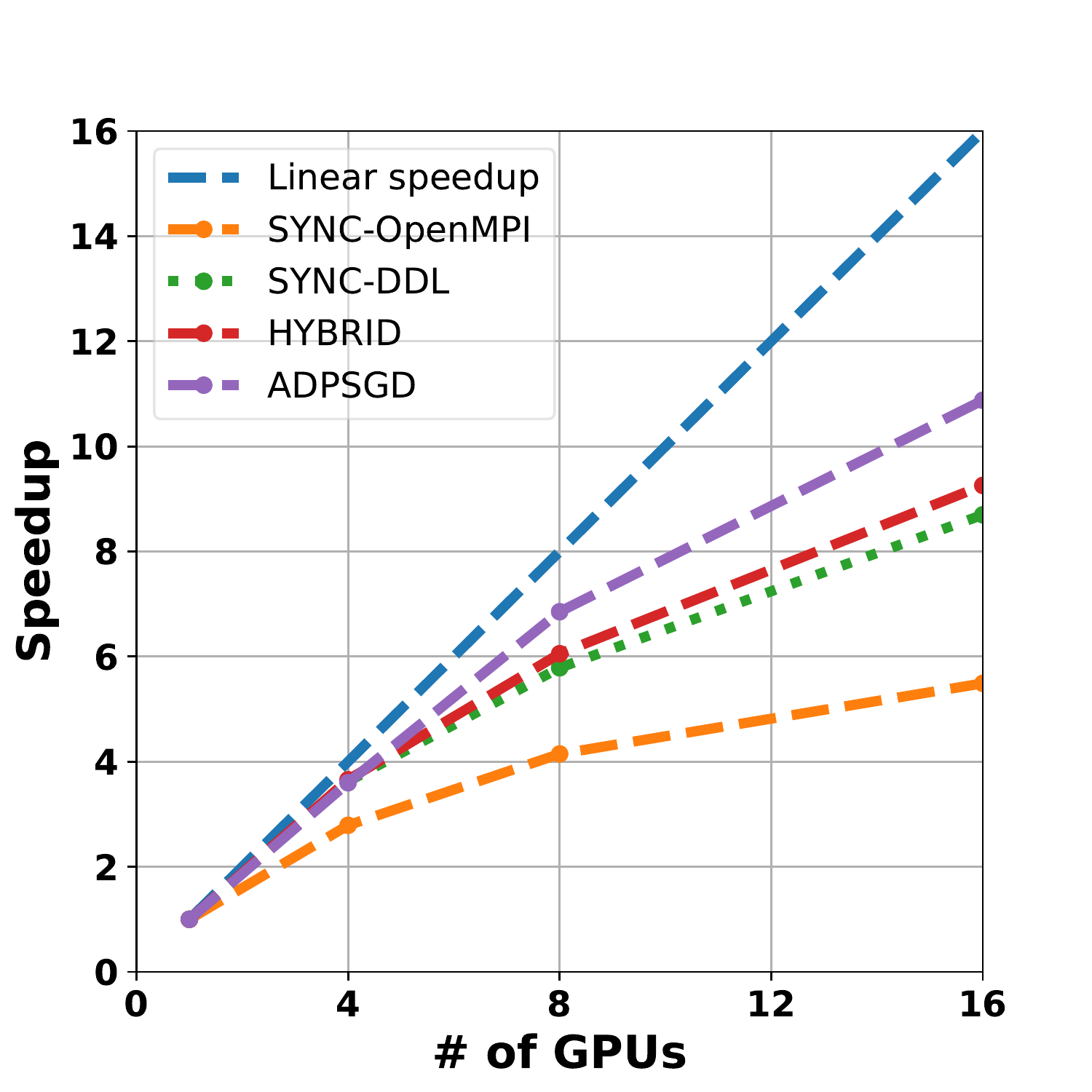}
    \label{fig:speedup}
  }
  \caption{Convergence/Speed-up comparison, 16 GPUs on \dyce.}
  \vspace{-0.5cm}
\end{figure}


\begin{table}[h]
\centering
\begin{tabular}{|l|l|l|l|l|}
\hline
         & Baseline & \sync & \hybrid & \adpsgd   \\ \hline
\swb     & 7.5\% & 7.6\% & 7.6\%   & 7.6\% \\\hline
\ch    & 13.0\% & 13.1\% & 13.1\%  & 13.2\%\\ \hline
\end{tabular}
\caption{WER comparison between baseline, \sync, \hybrid, and \adpsgd after training for 16 epochs.}
\label{tab:wer}
\end{table}
Our LSTM baseline trained on single GPU (batchsize 256) gives a WER of 7.5\%/13.0\% on the Switchboard/CallHome (SWB/CH) of the NIST Hub5 2000 evaluation test sets after the Cross-Entropy training, which is a competitive baseline. We compare this baseline with \sync, \hybrid, and \adpsgd in \Cref{fig:loss} for heldout loss and in \Cref{tab:wer} for WER.
\vspace{-0.3cm}

\subsection{Runtime Results}
\label{sec:results:runtime}

\Cref{fig:speedup} plots the speed-up for different algorithms up to 16 GPUs on \dyce. \sync-OpenMPI is found to be the slowest one. It is also found that \adpsgd achieves the best speed-up (about 11X over 16 GPUs) and finishes the training in 13.98 hours. \adpsgd did not achieve linear speed-up because it requires CPU-based weights averaging and GPU weights update to occur atomically which could be remedied by offloading GPU weights update to CPU. \hybrid does not outperform \sync-DDL significantly even though computation is long enough to hide the communication. This is because \hybrid asynchronously calls DDL which relies on NVIDIA NCCL library\cite{nccl} for the intra-server reduction. NCCL heavily competes with training for GPU resources (e.g., stream processor and memory) when used asynchronously.

\begin{figure}
    \includegraphics[width=\columnwidth, height=3cm]{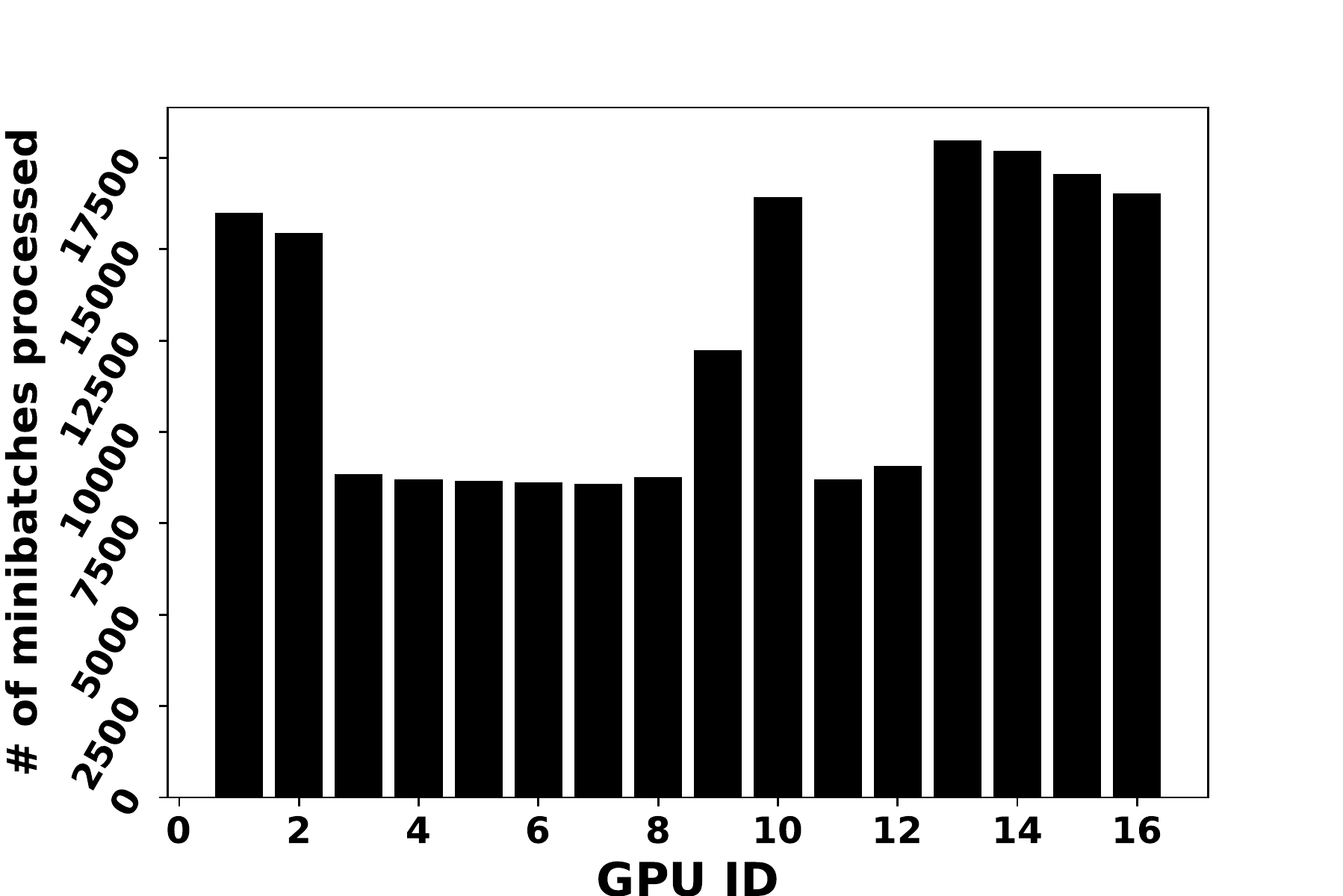}
    \caption{
       When half of \dyce is shared to run other tasks, \adpsgd balances the workload for different GPUs for the SWB2000-LSTM task.
    } \label{fig:lb}
  \vspace{-0.5cm}
\end{figure}

\Cref{tab:speedup_slow_node} shows the speed-up measured for each algorithm when one of the 16 GPUs slows down. \adpsgd is immune to the straggler problem, whereas the speedup of other algorithms quickly diminishes. \Cref{fig:lb} shows a snapshot of the number of minibatches processed by each GPU in one epoch when half of \dyce are shared by other users. \adpsgd automatically balanced the workload per GPU. \sync and \hybrid would enforce each GPU to process the same number of minibatches in this scenario. 

\begin{table}
        \centering

\resizebox{\linewidth}{!}{
        \begin{tabular}{|l|l|l|l|l|}
          \hline
          \multirow{2}{*}{\begin{tabular}[c]{@{}l@{}}Slowdown of\\ one GPU\end{tabular}} & \multicolumn{2}{l|}{\adpsgd} & \multicolumn{2}{l|}{\sync-DDL/\hybrid}         \\ \cline{2-5}
                                                                                          & Time/epoch (hr)    & Speed-up   & Time/epoch (hr) & Speed-up         \\ \hline
          no slowdown                                                                        & 0.87              & 10.88     & 1.09/1.03           & 8.70/9.26           \\ \hline
          2X                                                                                 & 0.89               & 10.63     & 1.67/1.63             & 5.71/5.83            \\ \hline
          10X                                                                                & 0.91               & 10.42     & 6.24/6.46             & 1.52/1.47            \\ \hline
          100X                                                                               & 0.92               & 10.38     & 57.73/60.80             & 0.16/0.16            \\ \hline

        \end{tabular}

}
\caption{Runtime and speedup when one GPU slows down by 2X-100X, 16 GPUs on \dyce. \adpsgd is immune to the straggler problem. }
        \label{tab:speedup_slow_node}
\vspace{-2mm}
      \end{table}

\subsection{Experiments on 32 GPUs}
We ran an experiment on \cs with 32 GPUs and batch size 80 per GPU\footnote{It takes 195 hours to finish training SWB2000 task on a V100 GPU, with a batchsize 80.}. \sync-DDL, \hybrid and \adpsgd complete one epoch in 0.75 hrs (16.25x speed-up), 0.71hrs (17.17x speed-up), and 0.83hrs (14.69x speed-up) respectively. \textit{{\hybrid trains SWB2000 to reach WER 7.6\% on SWB and WER 13.1\% on CH in 11.5 hrs.}}

\adpsgd requires more CPU resources than \sync and \hybrid to conduct weights averaging and passing weights. \cs has a less favorable CPU/GPU ratio. Furthermore, 8 GPUs share one PCIe bus on \cs and \adpsgd saturates the 16GB/s bandwidth quickly. \adpsgd would run significantly faster if deployed on clusters with higher CPU/GPU ratio, higher main memory bandwidth, and/or advanced GPU-CPU interconnect (e.g., NVLink\cite{nvlink}
).

%% file: future.tex
\section{Conclusion and Future work}
\label{sec:future}
In this paper, we made the following contributions: (1) we first described the hyper-parameter setup for SWB2000-LSTM speech recognition task using batch size of 2560, which is a sufficiently large batch size that enables efficient distributed training. (2) We implemented and compared different distributed learning algorithms for this task. Our system trains a model to WER 7.6\% on SWB and WER 13.1\% on CH in less than 12 hours. To the best of our knowledge, this is the fastest system that trains these tasks to this level of accuracy. Our future work includes: (1) Implement wait-free ADPSGD as proposed in \cite{adpsgd} to further improve convergence and runtime. (2) Experiment with hardware with better memory bandwidth, CPU/GPU ratio,  and CPU-GPU inter-connect. (3) Experiment with different types of speech recognition workloads. (4) Explore other methods to increase batch size and/or use mixed-precision training as in \cite{lars, tencent-imgnet}.

%% file: related.tex
\section{Related Work}
\label{sec:related}
Distributed DL have been applied to speech recognition\cite{deepspeech2,bmuf}, computer vision\cite{facebook-1hr}, language modeling\cite{nvidia-lm-scaling}, and machine translation\cite{fb-mt-scaling} tasks. To reduce the cost of communication, researchers have proposed gradient quantization\cite{msr-1bit,naigang-nips18} and gradient compression\cite{adacomp, terngrad}. All these works adopt a synchronous training method which would become unacceptably slow in a resource-sharing or Cloud environment\cite{gadei}. Asynchronous SGD, based on the parameter-server architecture, is known to have inferior performance and should be avoided when possible\cite{revisit-sync-sgd, facebook-1hr, deepspeech2, zhang2016icdm}. This work is the first that applies Asynchronous Decentralized Parallel SGD (ADPSGD), which has the theoretical guarantee to converge at the same rate as SGD\cite{adpsgd}, to the challenging SWB2000-LSTM task to achieve state-of-the-art model accuracy in a record time. 